\documentclass[%
 reprint,
 amsmath,amssymb,
 aps,showpacs
]{revtex4-1}

\usepackage{graphicx}
\usepackage{dcolumn}
\usepackage{bm}

\begin{document}

\preprint{APS/123-QED}

\title{Autler-Townes splitting via frequency upconversion \\ at ultra-low power levels in cold $^{87}$Rb atoms using an optical nanofiber } 

\author{Ravi Kumar$^{1,2}$}
\author{Vandna Gokhroo$^1$}
\author{Kieran Deasy$^1$}
\author{ S{\'i}le Nic Chormaic$^{1,}$}
\email{sile.nicchormaic@oist.jp}
\affiliation{$^1$Light-Matter Interactions Unit, Okinawa Institute of Science and Technology Graduate University, Onna, Okinawa 904-0495, Japan \\
$^2$Physics Department, University College Cork, Cork, Ireland
}%

\date{\today}

\begin{abstract}
The tight confinement of the evanescent light field  around the waist of an optical nanofiber makes it a suitable tool for studying nonlinear optics in atomic media. Here, we use an optical nanofiber embedded in a cloud of laser-cooled $^{87}$Rb for near-infrared frequency upconversion via a resonant two-photon  process.  Sub-nW powers of the two-photon beams, at 780 nm and 776 nm, co-propagate through the optical nanofiber and generation of 420 nm photons is observed.   A measurement of the Autler-Townes splitting provides a direct measurement of the Rabi frequency of the 780 nm transition. Through this method, dephasings of the system can be studied.  In this work, the optical nanofiber is used  as an excitation and detection tool simultaneously, and it highlights some of the advantages of using fully fibered systems for nonlinear optics with atoms.
\end{abstract}

\maketitle


Subwavelength diameter optical fibers, also known as optical nanofibers (ONFs), have recently emerged as a very useful tool for probing \cite{Nayak2007, Morrissey2009, Russell2012a}and trapping cold atoms \cite{Vetsch2010, Goban2012}, particularly due to the functionality of such nanofibers in the development of atom-photon hybrid quantum systems \cite{Hoffman2011, Beguin2014}. Aside from this research focus, ONFs have also been shown to be highly efficient tools for demonstrating nonlinear optics using very low light power levels in atomic systems \cite{Fleisch2005}.  More than a decade ago, Patnaik \textit{et al.} \cite{Patnaik2002} proposed a demonstration of slow light in an ONF surrounded by a nonlinear medium, such as atoms. More recently, two photon absorption by laser-cooled atoms using an ONF was proposed \cite{Laura2012}. Quantum interference effects, such as electromagnetically induced transparency (EIT) \cite{Spillane2008} and two-photon absorption  \cite{Hendrickson2010}, were  demonstrated using an ONF in rubidium (Rb) vapor, and nW level saturated absorption in a Xe gas was observed with an ONF \cite{Pittman2013}. This versatility of ONFs for nonlinear optics arises from the very high evanescent field intensities that can be achieved as a result of the very tight light confinement within a very small mode area over long distances of a few mm. For example, atoms surrounding an ONF experience an observable ac Stark shift on their hyperfine energy levels. Here, we study a two-photon excitation process, at 780 nm and 776 nm, in a cascade three-level configuration \cite{Ryan1993} in cold $^{87}$Rb atoms using an optical nanofiber.  We have observed frequency up-conversion for 776 nm probe power as low as 200 pW and Autler-Townes (A-T) splitting \cite{Autler1955} for $<$20 nW of 780 nm coupling power. These power levels are several orders of magnitude lower than those used in free space experiments \cite{Fox1993, Piotrowicz2011, Zhang2013}. The effect of varying the coupling power on the obtained A-T spectra is investigated and sources of dephasing within the system are considered.

\begin{figure}
\includegraphics[scale=0.5]{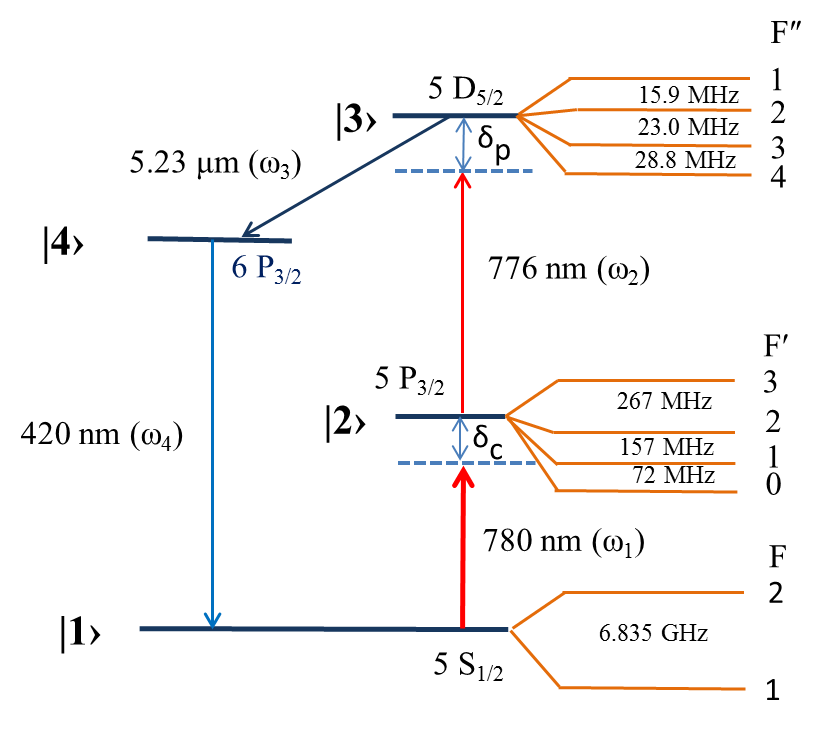}
\caption{\label{energy_level_diagram} Energy level diagram for $^{87}$Rb atoms showing the 780 nm coupling and 776 nm probe beams. }
\end{figure}

\begin{figure}
\includegraphics[scale=0.45]{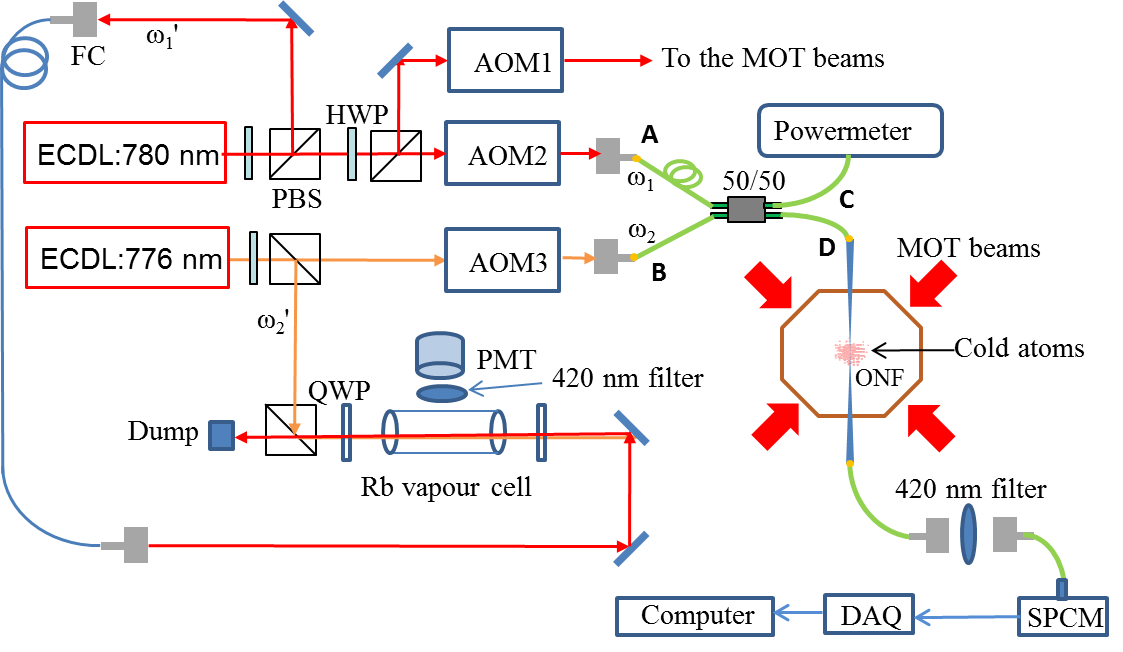}
\caption{\label{experimental_setup} Schematic of the experimental setup. 780 nm and 776 nm light are co-propagating in the nanofiber to generate blue photons from the cold atoms. An SPCM is used to detect the blue photons exiting from the nanofiber pigtail after filtering in free space. 780 nm and 776 nm light counter-propagating in a Rb vapor cell are used to obtain the reference signal for identification of the peaks in the two-photon spectrum. SPCM: Single photon counting module, ECDL: Extended cavity diode laser, AOM: Acousto-optic modulator, QWP: Quarter waveplate, HWP: Half waveplate, PMT: Photo multiplier tube, MOT: Magneto-optical trap, ONF: Optical nanofiber,  DAQ: Data acquisition card, FC: Fiber coupler.}
\end{figure}

We use laser-cooled $^{87}$Rb atoms, in a standard MOT configuration, the details of which are described elsewhere \cite{Kumar2015}.  We adopt a two-photon cascade three-level system where  5S$_{1/2}$ ($|1\rangle$) is the ground state, 5P$_{3/2}$ ($|2\rangle$) is the intermediate state, and 5D$_{5/2}$ ($|3\rangle$ is the excited state.  Relaxation of $|3\rangle$ via  6P$_{3/2}$ ($|4\rangle$) generates 420 nm blue light (Fig. \ref{energy_level_diagram}). A schematic of the experimental setup is shown in Fig. \ref{experimental_setup}. A Rb vapor cell is used to provide the reference frequencies for the two-photon transitions. A counter-propagating configuration is chosen since the linewidths are solely determined by the lifetime of the final state, $|3\rangle$ \cite{Bjorkholm1977}, and sharper peaks can be observed for reference purposes. The 420 nm blue fluorescence ($\omega_{4}$) generated in the vapor cell is monitored using a photomultiplier tube (PMT) with the aperture covered by a 420 nm filter (FWHM of 10 nm). We do not detect the infrared 5.23 $\mu$m ($\omega_{3}$) photons. All lasers used are extended cavity diodes (ECDL), one of which is locked to the 5S$_{1/2}$ F=2 $\rightarrow$ 5P$_{3/2}$ F$^\prime$ = 2 and F$^\prime$=3 crossover transition using a standard saturation absorption setup. This laser provides both the cooling beams for the MOT and the 780 nm coupling beam for the $|1\rangle$ $\rightarrow$ $|2\rangle$ transition (i.e., the 5S$_{1/2}$ F=2 $\rightarrow$ 5 P$_{3/2}$ F$^\prime$ = 3 transition). The second ECDL, at 776 nm, is scanned across the $|2\rangle$ $\rightarrow$ $|3\rangle$ probe transitions. A third ECDL (not shown in Fig. \ref{experimental_setup}) is used for the 780 nm repump beam in the MOT.

The ONF used  in the cold atom experiment is prepared from a commercial, single-mode optical fiber for 780 nm using a flame brushing technique  \cite{Ward2014}. It has a diameter of $\sim$350 nm, ensuring that only the fundamental mode propagates at 780 nm.  Note that higher modes may propagate in the ONF for 420 nm light. Transmission through the ONF at 780 nm is measured to be 84\%.  The ONF is mounted on an aluminium u-shaped mount and installed vertically in the vacuum chamber \cite{Morrissey2009}. The experiment is designed so that the cold atom cloud is centered on the waist of the ONF.  The atom cloud diameter is $\sim$0.8 mm and the  temperature is measured to be $\sim$200 $\mu$K using a time-of-flight technique.  Alignment of the atom cloud and the ONF is optimized via two small magnetic shim coils that are used to overlap the fiber with the densest part of the cloud.  This is done while monitoring the spontaneous emission from the atoms coupling into the ONF.  The cloud contains $\sim$10$^7$ atoms; however, if we consider the evanescent field decay length for 780 nm light, there are typically $<$10 atoms in the evanescent field region, and the photon signals collected via the ONF can be considered to be directly related to emissions from such low atom numbers.

The 780 nm and 776 nm beams used in the vapor cell reference measurements ($\omega_{1}^{\prime}$ and $\omega_{2}^{\prime}$, respectively) are split in order to obtain the required frequencies ($\omega_1$ and $\omega_2$, respectively) for two-photon excitation in the cold atoms via the ONF.  The 780 nm beam from the ECDL is double-passed through AOM2 (Fig. \ref{experimental_setup}) using  the `+1' order to obtain $\omega_1$, which is 14 MHz red-detuned from the 5S$_{1/2}$ F=2 to 5P$_{3/2}$ F$^\prime$=3 cooling transition. The 776 nm beam from the second ECDL is double-passed through AOM3 using the `-1' order to ensure that $\omega_{1}^{\prime}$ + $\omega_{2}^{\prime}$ $=$ $\omega_1$ + $\omega_2$. This permits us to directly compare the spectra obtained from the cold atoms and the vapor cell in real time. Circular polarization of the same handedness is used for all 780 nm and 776 nm beams.

\begin{figure}
\includegraphics[scale=0.3]{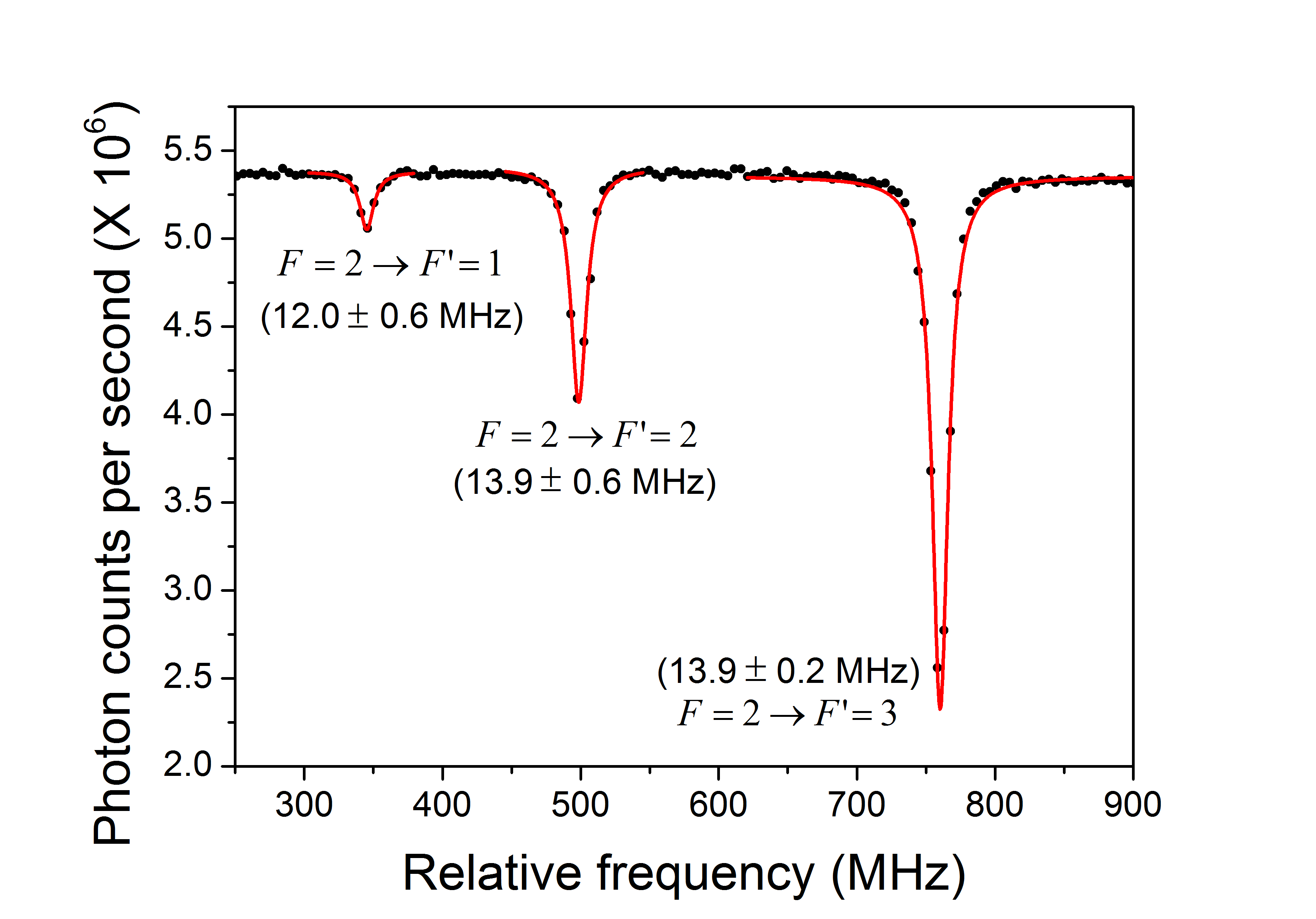}
\caption{\label{1-photon_absorption} Transmission of 780 nm light passing through the ONF with cold $^{87}$Rb atoms around the waist. The laser is scanning  across the 5S$_{1/2}$ F=2 to 5P$_{3/2}$  possible transitions. This spectrum is obtained using an additional ECDL laser in scan mode. The data are fit (red curves) to Lorentzian profiles and the obtained linewidths for the relevant transitions are indicated at the peaks. }
\end{figure}

$\omega_1$  is sent through port A of a 50:50 fiber beam splitter, while $\omega_2$ passes through port B (Fig. \ref{experimental_setup}). In order to excite the cold $^{87}$Rb atoms from  $|1\rangle$ to $|3\rangle$ in the two-photon cascade system via the ONF, one output port, D,  of the fiber splitter is spliced to one pigtail of the nanofiber ((Fig. \ref{experimental_setup}).  The other output port (C) is connected to a power meter to monitor beam powers. The measured power is proportional to the power at the nanofiber waist, any differences being due to the ONF transmission losses at a particular wavelength. Hence, if we assume equal losses at both sides of the taper, the measured power can be taken as 1.1 times the waist power for both 780 nm and 776 nm wavelengths. A typical transmission spectrum of the 780 nm light (for input power of 1.8 nW) through the ONF is shown in Fig. \ref{1-photon_absorption} with no $\omega_2$ present. As the laser is scanned across the 5S$_{1/2}$ F=2 to 5P$_{3/2}$ F$^\prime$ transitions, absorption dips appear. If we add $\omega_2$  into the nanofiber, 420 nm (blue) photons are generated within the atom cloud via four-wave mixing for co-propagating coupling and probe beams \cite{Kienlen2013}.  $\omega_1$ and $\omega_2$ excite the atoms from $\left |1 \right\rangle $ to  $\left |3 \right\rangle $    via  $\left |2 \right\rangle $.  In the relaxation process from $\left |3 \right\rangle $ to $\left | 4 \right\rangle $  and from $\left | 4 \right\rangle $ to $\left |1 \right\rangle $ $ \omega_3 $ ($\vec{k}_{IR}$) and $ \omega_4 $ ($\vec{k}_{blue}$) are generated, respectively. The decay probability from $\left |3 \right\rangle $ to $\left |4 \right\rangle $ is 35$\%$ and from $\left | 4 \right\rangle $ to $\left |1 \right\rangle $ is 31$\%$ \cite{HEAVENS1961}. The four frequencies are related by the frequency-matching condition to satisfy conservation of energy, $ \omega_1$ + $\omega_2$  = $\omega_3$ + $\omega_4 $, whereas momentum conservation requires the phase-matching relation, $\vec{k}_{780}$ + $\vec{k}_{776}$ = $\vec{k}$$_{IR}$ + $\vec{k}$$_{blue}$ to be satisfied \cite{Akulshin2009}. In this system, the phase-matching condition must be satisfied since $\omega_1$ and $\omega_2$ co-propagate and the blue light, $\omega_4$, must be produced in the forward direction. However, we did not try to observe the presence of blue photons in the backward direction due to constraints in the experimental setup. The blue photons couple into the nanofiber and propagate along it. The guided light is coupled out of the ONF and passed through a 420 nm (FWHM : 10 nm) filter before reaching the single photon counter (SPCM).   The filter serves to eliminate any residual excitation beams, or other 780 nm photons, coupled to the nanofiber from the atom cloud or the MOT beams. Detection of blue photons serves as a signature of the two-photon absorption process in the evanescent field region, hence the ONF acts as both the excitation and detection tool simultaneously.

\begin{figure}
\includegraphics[scale=0.7]{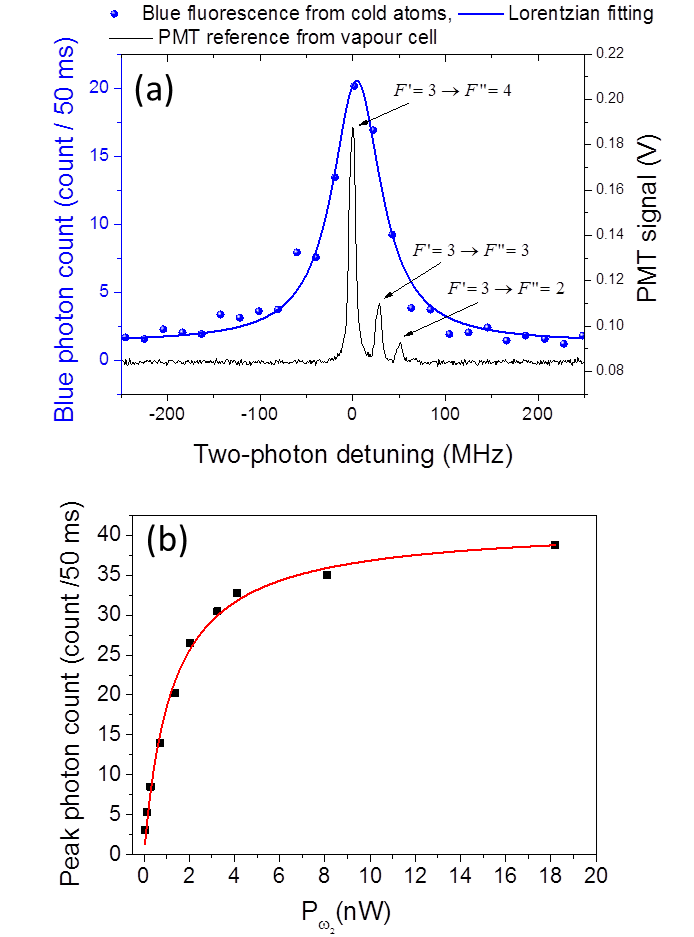}
\caption{\label{blue_PMT}(a) 420 nm photon count rate (blue dots) as $\omega_{2}$ is scanned across the 5P$_{3/2}$ to 5D$_{5/2}$ transition.  An intermediate  power (1.4 nW) of $\omega_{2}$ is input into the nanofiber, whereas any 780 nm present is only from the MOT beams. The data are fitted (solid blue curve) to a Lorentzian profile. The black spectrum is the corresponding reference signal obtained from the vapor cell when the 780 nm and 776 nm beams are counter-propagating. The hyperfine transitions associated with each peak are indicated. (b) Maximum blue photon count (i.e. the peak value of the curve in (a)) for different powers of $\omega_{2}$.  Here, the 780 nm contribution also only from the MOT beams. The  red curve is a theoretical fit to yield the saturation power as 1.24$\pm$0.12 nW.  This corresponds to $\sim$1.1 nW of 776 nm power at the waist. }
\end{figure}

To study the influence of coupling and probe power on the two photon process we start with no $\omega_1$ light through port A of the 50:50 splitter (Fig.\ref{experimental_setup}).  $\omega_2$  is sent through port B and we observe blue emission from the cold atom cloud.  Hence, we deduce that excitation from $\left |1 \right\rangle $ to $\left |2 \right\rangle $ purely from the MOT beams is sufficient to initiate the two-photon process, which we observe for as little as 200 pW of power in $\omega_2$. Fig. \ref{blue_PMT}(a) shows the typical blue counts detected on the SPCM when $\omega_2$ was scanned across the two-photon transition.  The peak in the observed spectrum occurs at the same two-photon frequency detuning as that which gives rise to the strongest observed transition in the vapor cell. When we plot the peak blue photon count rate as a function of probe power in $\omega_2$ we see there is saturation behavior even though we operate in the nW region (Fig. \ref{blue_PMT}(b)).

The linewidth obtained for the two-photon spectrum from cold atoms Fig. \ref{blue_PMT}(a) is broader than the natural linewidths of the intermediate ($\sim$5.9 MHz) and final levels ($\sim$0.66 MHz). This could arise from dephasing introduced to both the levels due to the presence of MOT beams at all times during measurements. Power broadening and the ac Stark effect from the MOT beams would give partial broadening. The other contributions in the broadening may come from the presence of the 5D$_{5/2}$ state manifold and atom-fiber surface interactions \cite{Lekien2007, Minogin2009}. Note that there is not much observable broadening when we only use a 780 nm probe beam for standard one-photon absorption (Fig. \ref{1-photon_absorption}). This may be due to the effect of light-induced dipole forces on the atomic cloud \cite{Sague2007}. In our case, we measure $\sim$14 MHz linewidth even when using nW of power.

\begin{figure}
\includegraphics[scale=0.35]{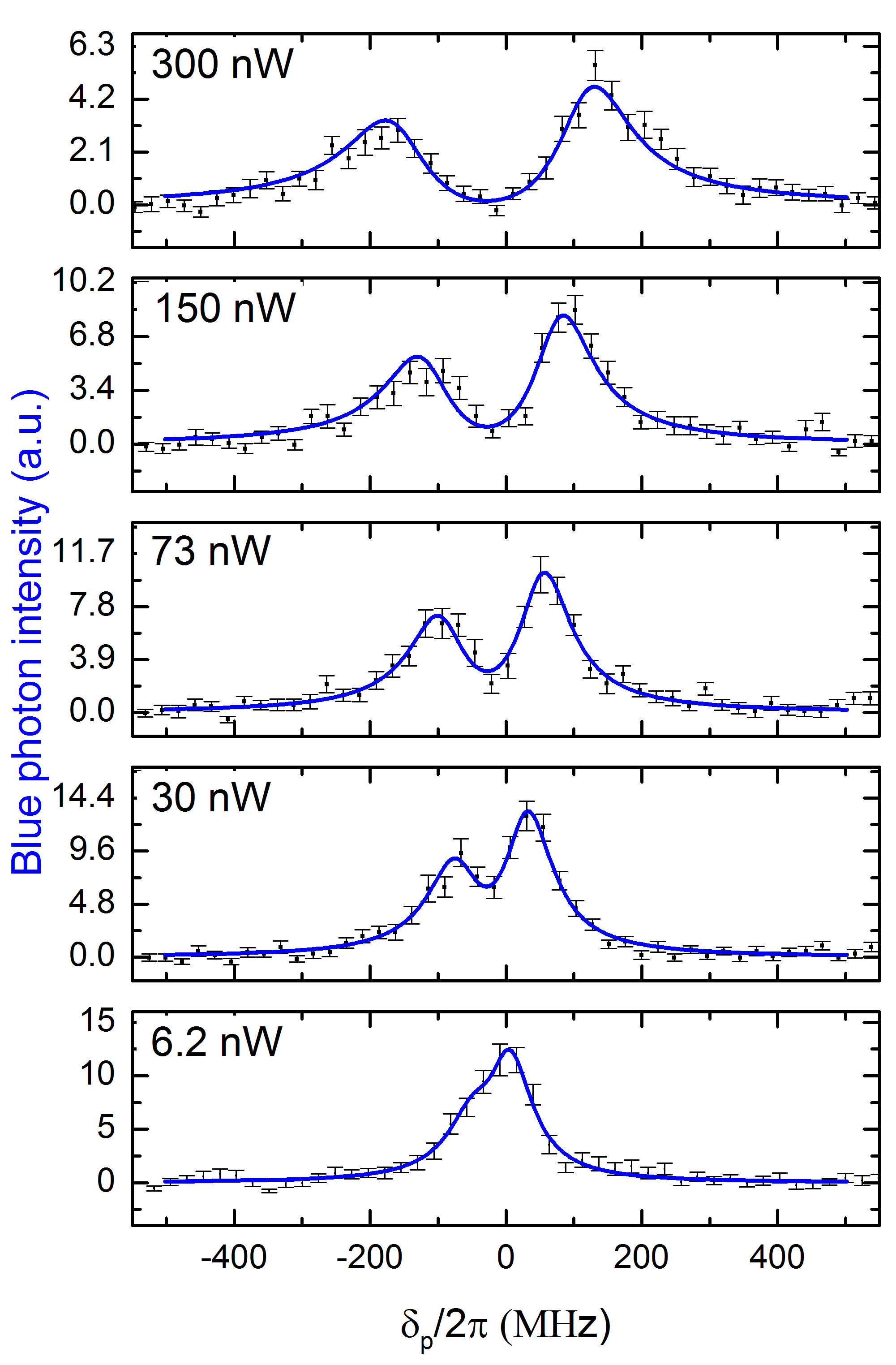}
\caption{\label{Splitting} Blue fluorescence from the atoms collected via the ONF for different powers in $\omega_1$, which is 14 MHz red-detuned from the 5S$_{1/2}$ F=2 to 5P$_{3/2}$ F$^\prime$=3 transition,  while $\omega_2$ is scanned across the 5P$_{3/2}$ F$^\prime$=3  to 5D$_{5/2}$ hyperfine levels. The power for $\omega_2$ is fixed at 0.5 nW.  $\delta_p$ is the detuning of $\omega_2$ as indicated in Fig. \ref{energy_level_diagram}.  $\omega_1$ is held at the same frequency as the cooling beams. Asymmetry in the observed A-T doublet is due to the fact that $\omega_1$ is not on resonance.  Solid lines are theoretical fits to the data using Eq.(1).}
\end{figure}

\begin{figure}
\includegraphics[scale=0.3]{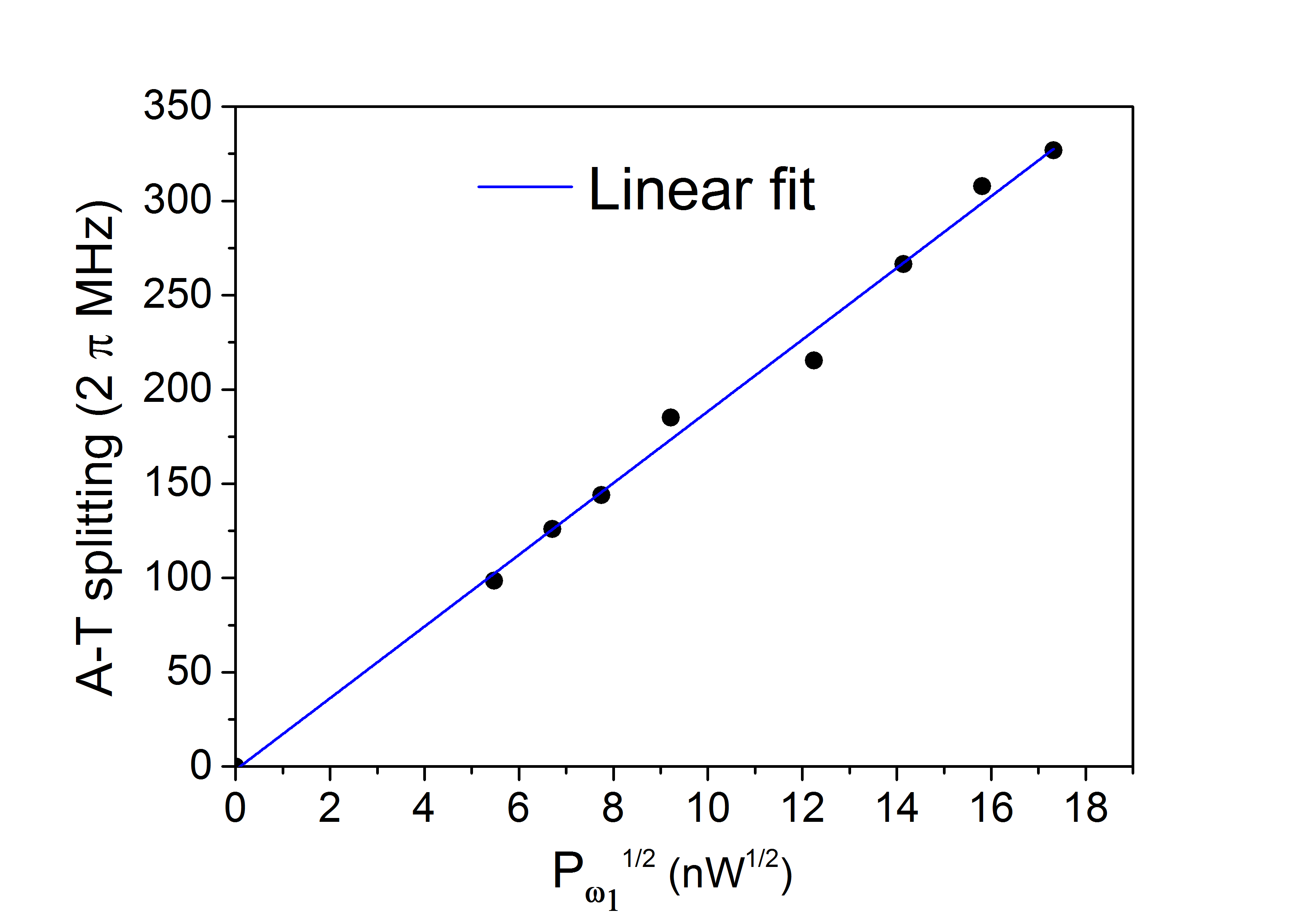}
\caption{\label{A-T_splitting_vs_Power} Measured A-T splitting as a function of the square-root of the power in the 780 nm coupling beam.}
\end{figure}

Next, in order to study the effect of the very strong evanescent field intensities on atomic transitions, we introduce the coupling laser, $\omega_1$, into the ONF via port A of the fiber coupler.  The power in the coupling beam, $P_{\omega_1}$, is varied while the probe power, $P_{\omega_2}$, is fixed at 500 pW.  This value was chosen to ensure that sufficient 420 nm photons are obtained for detection. We observe that the peak blue photon count increases with $P_{\omega_1}$ and the width of the spectrum broadens (data not shown here). For a  $P_{\omega_1}$ $\sim$ 20nW, the obtained spectrum clearly splits into two peaks. The peak separation increases as  $P_{\omega_1}$ increases (Fig. \ref{Splitting}). This is known as Autler-Townes (A-T) splitting and is caused by the ac Stark effect of the 780 nm transition in the presence of a strong coupling beam (\cite{Autler1955}). The A-T splitting is plotted for different values of $P_{\omega_1}$ (Fig. \ref{A-T_splitting_vs_Power}(a)  and we see that it is directly proportional to the square-root of $P_{\omega_1}$ as expected. The A-T splitting spectrum is given by the imaginary part of the density matrix term $\rho_{23}$ \cite{Zhang2013}

\begin{equation}
Im(\rho_{23}) \varpropto  \frac{4(\delta_p + \delta_c)^2 \gamma_2+\gamma_3(|\Omega_c|^2+ \gamma_2\gamma_3)}{||\Omega_c|^2-(i\gamma_3-2\delta_p)[i\gamma_2-2(\delta_p+\delta_c)]|^2},
 \end{equation}

where $\gamma_{2(3)}$, $\delta_{c(p)}$, and $\Omega_{c}$ represent dephasing, detuning and the Rabi frequency of coupling transition respectively.

\begin{figure}
\includegraphics[scale=0.25]{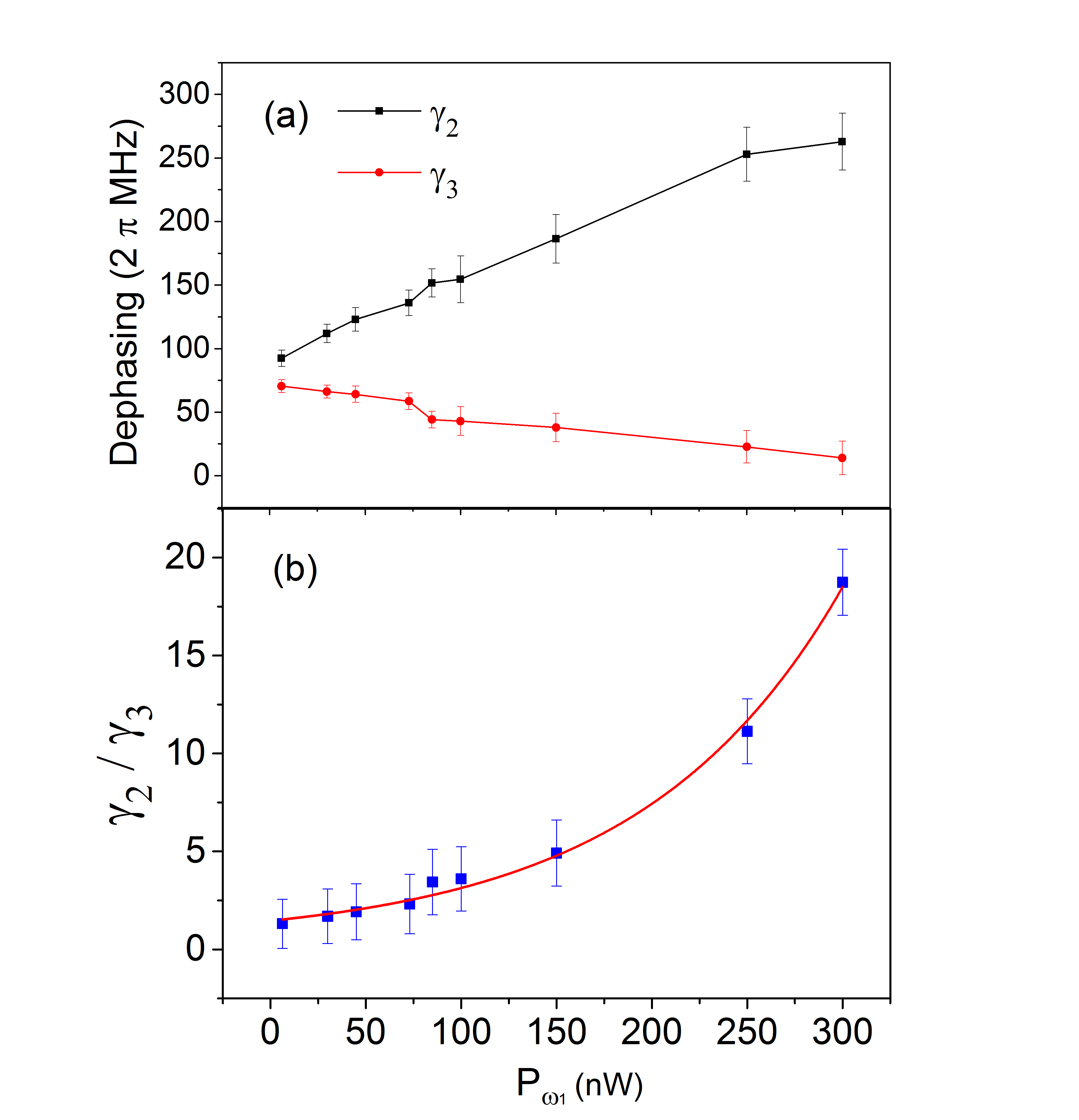}
\caption{\label{dephasingRatio} (a) Variation of $\gamma_{2}$ and $ \gamma_{3}$ as a function of coupling power,$P_{\omega_1}$. Solid lines are guides to the eye. (b)Ratio of the two dephasing terms, $\gamma_{2}/ \gamma_{3}$, for different coupling powers, $P_{\omega_1}$, solid line is an exponential fit.}
\end{figure}

 Fitting Eq. (1) to the experimental data (Fig. \ref{Splitting}), we can obtain values for $ \gamma_{2}$, $ \gamma_{3}$ and $\Omega_{c}$. The dephasing terms, $ \gamma_{2}$ and $ \gamma_{3}$, are found to increase and decrease, respectively, with an increase in $P_{\omega_1}$.  If we consider the ratio of the dephasings we see that $\gamma_{2}/ \gamma_{3}$ follows an exponential increase as a function of $P_{\omega_1}$ (Fig. \ref{dephasingRatio}(b)).  This effect could be due to the heating influence of $\omega_1$ on the atoms. As the power in $\omega_1$ is increased, the atom cloud temperature should increase, thereby leading to a change in atom cloud density and a possible increase in atom-atom or atom-surface interactions. From our observations, the influence of heating on atoms in the lower excited state is more significant than on those in the upper excited state. This is in stark contrast to that observed for highly excited states in ultracold cesium, where the dephasing due to  strong interactions between Rydberg atoms is considered to be larger than other dephasing terms in the system \cite{Zhang2013}. In order to gain better insight on the dynamics at play, a thorough study of the $\gamma_{2}/ \gamma_{3}$ ratio as a function of detuning of $\omega_1$, atom temperature, and cloud density should be conducted.

In conclusion, we have observed frequency up-conversion and A-T splitting for ultra-low power levels (nW) in an atom+nanofiber system.  The splitting is observed for ultra-low powers of the coupling field in the evanescent  region of the nanofiber. If we consider 50 nW of coupling power propagating in the ONF, we can assume that there are typically less than two photons in the interaction volume at any given time \cite{Tong2004, Hendrickson2010}. However, such power levels are used frequently in nanofiber experiments and  it is important to take into account any induced shifts in the energy levels that may arise \cite{Lee2014}. In the high intensity regime, the Rabi frequency for the coupling transition is approximately equal to the A-T splitting \cite{Stenholm1984} and this method allows us to measure it directly for an atom+nanofiber system. Otherwise, due to the difficulties in exactly determining  nanofiber parameters such as the influence of fiber surface on energy levels, the effective position of the atoms in the evanescent field, waist size of the ONF etc., this could be challenging to estimate with any accuracy. The observation of  nonlinear phenomena using an optical nanofiber in a cold atom system increases the versatility of such devices and may be useful for demonstrations of single photon all-optical switching \cite{Hendrickson2013}, or quantum logic gates \cite{Krishnamurthy2013} at ultra-low powers. The efficiency of the process may be improved by optimizing the beam polarizations \cite{Vernier2010} at the nanofiber waist, a technique that relies on optimum control on light propagation in ultrathin fibers \cite{Ravets2013, Mitsch2014, Kumar2015}.


RK and VG contributed equally to this work. This work was supported by the Okinawa Institute of Science and Technology Graduate University. S.N.C. is grateful to JSPS for partial support from Grant-in-Aid for Scientific Research (Grant No. 26400422). The authors would like to thank E. Brion, M. Lepers, and R. Gu\'{e}rot for useful discussions.





\end{document}